\documentclass[prc,showpacs,superscriptaddress,floatfix]{revtex4}
\usepackage{graphicx,amssymb,amsmath}

\newlength{\ldag}
\newcommand{\crs}{s^\dagger}
\newcommand{\ans}{{s^{\phantom\dagger}\hspace{-\ldag}}}

\newcommand{\crL}{L^\dagger}
\newcommand{\anL}{{L^{\phantom\dagger}\hspace{-\ldag}}}
\newcommand{\anLt}{{\tilde{L}^{\phantom\dagger}\hspace{-\ldag}}}

\newcommand{\crb}{b^\dagger}
\newcommand{\anb}{{b^{\phantom\dagger}\hspace{-\ldag}}}

\newcommand{\crc}{c^\dagger}
\newcommand{\anc}{{c^{\phantom\dagger}\hspace{-\ldag}}}

\newcommand{\crx}{\xi^\dagger}

\newcommand{\crP}{P^\dagger}
\newcommand{\anP}{{P^{\phantom\dagger}\hspace{-\ldag}}}

\begin{document}

\title{Two-level interacting boson models beyond the mean field}

\author{Jos\'e M. Arias}
\affiliation{Departamento de F\'{\i}sica At\'omica, Molecular y
Nuclear, Facultad de F\'{\i}sica, Universidad de Sevilla,
Apartado~1065, 41080 Sevilla, Spain}

\author{Jorge Dukelsky}
\affiliation{Instituto de Estructura de la Materia, CSIC, Serrano
123, 28006 Madrid, Spain}

\author{Jos\'e Enrique Garc\'{\i}a-Ramos}
\affiliation{Departamento de F\'{\i}sica Aplicada, Universidad de
Huelva, 21071 Huelva, Spain}

\author{Julien Vidal}
\affiliation{Laboratoire de Physique Th\'eorique de la Mati\`ere
Condens\'ee, CNRS UMR 7600,
Universit\'e Pierre et Marie Curie, 4 Place Jussieu,
75252 Paris Cedex 05, France}

\begin{abstract}
The phase diagram of two-level boson Hamiltonians, including the
Interacting Boson Model (IBM), is studied beyond the standard mean
field approximation using the Holstein-Primakoff mapping. The limitations of the usual intrinsic state (mean field) formalism concerning finite-size effects are pointed out. The analytic results  are compared to numerics obtained from exact diagonalizations. Excitation energies and occupation numbers are studied in different model space regions (Casten triangle for IBM) and especially at the
critical points.
\end{abstract}


\maketitle

\section{Introduction}

The concepts of phase transition and critical points are defined,
strictly speaking for macroscopic systems. However, it has been
recently suggested that precursors of phase transitions can be
observed in finite-size mesoscopic systems \cite{Iachello04}.  In
Nuclear Physics the different nuclear shapes and the phase transitions
between them are conveniently studied within the Interacting Boson
Model (IBM) \cite{Iachello87}. This was early recognized after the
introduction of the model \cite{Dieperink80_1,Feng81,Dukelsky84,
Frank89,Lopez96} but has been studied thoroughly in the last few years
\cite{Cejnar00,Jolie02,Cejnar03,Arias03_1,Arias03_2,
Rowe04_1,Turner05,Rosensteel05,Rowe05,Cejnar05,Heinze06} after the
introduction of the concept of critical point symmetries
\cite{Iachello00,Iachello01,Iachello03}.  Since the IBM is formulated
from the beginning in terms of creation and annihilation boson
operators, its geometric interpretation in terms of shape variables is
usually done by introducing a boson condensate with two shape
parameters, $\beta$ and $\gamma$ (order parameters) \cite{Ginocchio80,
Dieperink80_1}.  The parameter $\beta$ is related to the axial
deformation of the system, while $\gamma$ measures the deviation from
axial symmetry. The equilibrium shape of the system is obtained by
minimizing the expectation value of the Hamiltonian in the intrinsic
state. Shape phase transitions are studied theoretically using one (or
a few) control parameter(s) in the Hamiltonian. These control parameters
drive the system in different phases characterized by order parameters
and allows one to study in a simple way phase transitions and
critical points in Nuclear Physics.

The phase diagram of the IBM has been studied with several approaches
\cite{Cejnar00,Jolie02,Cejnar03,Arias03_1,Arias03_2,Turner05,Rowe05,Cejnar05,Heinze06}
and it is well known that the dynamical symmetry associated to $U(5)$
corresponds to a spherical shape ($\beta=0$), the dynamical symmetry
$SU(3)$ is associated to an axially deformed shape ($\gamma=0,\pi/3,
~\beta \neq 0$) and the dynamical symmetry $O(6)$ is related to a
$\gamma$-unstable deformed shape ($\beta \neq 0$ and
$\gamma$-independent). These symmetry limits are usually represented
as the vertices of a triangle (Casten triangle) \cite{Casten90}. Phase
transitions between these shapes have been widely studied and it is
known that the phase transition from $U(5)$ to $O(6)$ is second order
while any other transition within the Casten triangle from a spherical
to a deformed shape is first order \cite{Jolie02,Arias03_2}. These
studies have been performed, as mentioned above, by using the
intrinsic state formalism.  However, it is known that this approximate
method is only correct at leading order in a $1/N$ expansion where $N$
is the number of bosons. In this paper, we present a method to go
beyond this order and compute finite-size corrections to several
spectroscopic observables. We stress that $1/N$ corrections obtained
with the intrinsic state formalism (or Hartree-Bose method) are in
general incorrect and give no information on the proper finite-size
corrections.

The paper is organized as follows: first the model Hamiltonian is
introduced in Sect. II. In Sect. III, the Holstein-Primakoff mapping
\cite{Holstein40} is performed leading to a boson Hamiltonian in which
we retain terms in orders $N$, $N^{1/2}$ and $N^0$. Then a Bogoliubov
transformation is performed to diagonalize the Hamiltonian and to
study both the symmetric (spherical) and the broken (deformed) phases.
All this is done in general for two-level boson models in which the
lowest level is a scalar ($s$) boson while the upper level is an
arbitrary $L$ boson. The IBM corresponds to the particular case $L=2$
($d_\mu$ bosons). In addition to the IBM, we present results for the
case $L=0$ as an illustration of the general method. In Sect. IV we
compare the analytical results with exact numerical diagonalizations
for different paths along the Casten triangle.  Finally, Sect. V is
for the summary and conclusions.
%
%
\section{The Model}
%
%
As it has been noted before \cite{Turner05}, the experimental
exploration of the shape transition and critical points in nuclei is
difficult due to the lack of a continuous control parameter.  However,
in theoretical studies this limitation is overcome by using a
Hamiltonian written in terms of one or more control parameters that
can vary continuously. In this work, we consider a two-level boson
model in which the lowest level is characterized by a zero angular
momentum ($s$-boson) while the upper level has an arbitrary angular
momentum $L$.  The Hamiltonian proposed is a generalization of the IBM
consistent-Q formalism (CQF) \cite{Warner82}, which depends on two
control parameters $x$ and $\chi$
\begin{equation}
\label{eq:hamiltonian}
  H=x  \: n_L-\frac{1-x}{N}  Q^{\chi}\cdot Q^{\chi},
\end{equation}
where $n_L= \sum_\mu L^\dag_\mu L_{\mu}$ is the operator for the
number of bosons in the upper level, $N$ is the total number of
bosons, the symbol $\cdot$ stands for the scalar product defined as
$a\cdot b=\sum_{\mu=-L}^{+L} (-1)^\mu a_\mu b_{-\mu}$, and $Q^{\chi}$
is a multipole operator written as,
\begin{equation}
  \label{eq:Q}
 Q_\mu^{\chi}=(\crs \anLt+\crL \ans)^{(L)}_\mu+\chi [ \crL
\times \anLt ]^{(L)}_\mu ,
\end{equation}
where $\anLt_\mu=(-1)^\mu \anL_{-\mu}$.
For $L=2$, ($d$-bosons) the Hamiltonian (\ref{eq:hamiltonian}) is the
well-known CQF Hamiltonian for IBM. Though it is not the most general
IBM Hamiltonian, it captures the most important low energy properties
of a wide range of nuclei \cite{Warner83,Chou97,Fossion02}.  In
particular, it is general enough to describe different nuclear phases
and quantum phase transitions, and it has been used for that purpose
at the mean field level \cite{Jolie02,Arias03_2,Cejnar03}.

The Hamiltonian (\ref{eq:hamiltonian}) comprises different models
depending on the value of $L$. For instance, for $L=1$ the Hamiltonian
is appropriate for studying the phase diagram of the vibron model
\cite{Iachello95} of interest in Molecular Physics.

\section{Mean field and beyond}

The usual way of getting the phase diagram of the model
(\ref{eq:hamiltonian}) is to introduce shape variables. This can be
done by considering the intrinsic state formalism, also called
Hartree-Bose approximation, \cite{Ginocchio80,Dieperink80_1,
Dieperink80_2,Dukelsky84}. In this approach, the ground state is a
variational state built out of a condensate of ``dressed'' bosons,
that are independent bosons moving in the average nuclear field. For
$L=2$, these bosons are defined as
\begin{equation}
\label{bc}
\Gamma^\dagger_c = \frac{1}{\sqrt{1+\beta^2}} \left (s^\dagger + \beta
\cos     \gamma          \,d^\dagger_0          +\frac{1}{\sqrt{2}}\beta
\sin\gamma\,(d^\dagger_2+d^\dagger_{-2}) \right),
\end{equation}
and the $N$ boson condensate is,
\begin{equation}
\label{condensate}
| c \rangle = \frac{1}{\sqrt{N!}} (\Gamma^\dagger_c)^N | 0 \rangle.
\end{equation}
The variational variables $\beta$ and $\gamma$ are the order
parameters of the system and their equilibrium values are fixed by
minimizing the expectation value of the energy. The expression of this
energy can be found in many references
\cite{Ginocchio80,Dieperink80_1, Dieperink80_2,Garcia98} and can be
written schematically as follows,
\begin{equation}
\label{eq:gsener}
E(N,~\beta,~\gamma,~x,~\chi)=N F^{(1)}(N,~\beta,~\gamma,~x,~\chi) +
(N-1) F^{(2)}(N,~\beta,~\gamma,~x,~\chi),
\end{equation}
where $F^{(1)}(N,~\beta,~\gamma,~x,~\chi)$ is the matrix element of
the one-body operators divided by $N$ and
$F^{(2)}(N,~\beta,~\gamma,~x,~\chi)$ is the matrix element of the
two-body operators divided by $N-1$.  Note that there is no $N^2$
dependence in the two-body operator due to the definition of the
Hamiltonian. Actually, the only relevant contribution is the leading
one (order $N$) since the next one ($N^0$ for instance) are incomplete
as explained below.

%

For the standard IBM Hamiltonian ($L=2$), with an attractive
quadrupole interaction, the nucleus always becomes axially deformed,
either prolate ($\gamma=0$) for $\chi<0$ or oblate ($\gamma=\pi/3$)
for $\chi>0$. As a consequence, the parameter $\gamma$ can be
incorporated in the value of $\beta$.  $\beta>0$ corresponds to
$\gamma=0$ while negative $\beta$ implies $\gamma=\pi/3$. In the case
$\chi=0$ the nucleus becomes $\gamma$ unstable, {\it i. e.}, the
energy is independent of $\gamma$.

In this framework, one-phonon excitations above the ground state are
constructed by directly replacing in the ground state (\ref{condensate})
a condesate boson
by an excited boson (TDA method) or by including ground state
fluctuations (RPA method) \cite{Dukelsky84,Leviatan87,Garcia98}.  For
$L=2$, there are five excited phonons that are characterized by their
angular momentum projection $K$ and can be labelled as:
$\beta$-excitation with $K=0$, $\gamma$-excitations with $K=\pm 2$ and
finally two $K=\pm 1$ excitations. It should be noted that not all the 
excited phonons
are always physical, some of them become spurious, Goldstone bosons
associated with broken symmetries. This is the case for 
axially deformed nuclei, the $K=\pm 1$ excitations are Goldstone, 
spurious
bosons because the state constructed with this excitation corresponds
to a $O(3)$ rotation of the whole system. In the case of $\gamma$
unstable nuclei, the $K=\pm 2$ excitations also become Goldstone
bosons and are related with $O(5)$ rotations of the ground state. In
the case of $L=0$ only a $K=0$ excitation exists and it is directly
related, as we will see, with the $\beta$ band of the IBM
\cite{Vidal06_1}.

The mean field description of the ground state energy just mentioned
is only valid at order $N$. The first quantum corrections can be
obtained within the RPA formalism.  Alternatively, the
Holstein-Primakoff expansion \cite{Holstein40} offers a simple and
natural expansion in powers of $1/N$. The advantages of this
transformation are: it is Hermitian, preserves the boson commutation
relation, provides a correct expansion in powers of $N$ and its
leading order coincides with the mean field contribution.

The Holstein-Primakoff expansion consist in eliminating the $s$-boson
transforming the bilinear boson operators in the following way,
%
%
\begin{eqnarray}
\label{eq:def1}
L^\dagger_\mu L_\nu &=& \crb_\mu \anb_\nu,\\
L^\dagger_\mu \ans &=& N^{1/2} \crb_\mu (1-n_b/N)^{1/2}=(\crs L_\mu)^\dag,\\
\crs \ans &=& N-n_b,
\label{eq:def3}
\end{eqnarray}
%
%
where the  $b$-bosons satisfy $[\anb_\mu,\crb_\nu]=\delta_{\mu,\nu}$. The mapping fulfils the commutation relations at each order in $N$ in the Taylor expansion of the square root.

We next introduce the $c$-bosons through a shift transformation
%
\begin{equation}
\label{eq:shift}
  \crb_\mu = \sqrt{N} \lambda^*_\mu + \crc_\mu,
\end{equation}
%
where the $\lambda_\mu$'s are complex numbers which form a
$(2L+1)$-dimensional vector. This shift allows for a macroscopic occupation number $n_b$. Thus, it  allows to consider at the same time the spherical, setting $\lambda_\mu=0$ for all $\mu$, and the deformed phase,
$\lambda_\mu\neq 0$. In this latter situation, we shall only consider the case  $\lambda_0\neq 0$ without loss of generality. 
The Hamiltonian then reads,
%
%
%
%
\begin{eqnarray}
  H &=& N^1 \lambda_0^2\left\{ 5x-4-4(x-1) \lambda_0^2 +(x-1)\chi
    \alpha^{(L)}_{0,0} \lambda_0 \left[4  (1-\lambda_0^2)^{1/2} +
      \chi  \alpha^{(L)}_{0,0} \lambda_0 \right]\right\} + \nonumber\\
  &&N^{1/2}\lambda_0\left(\crc_0+\anc_0 \right)
  \left\{ 5x-4-8\lambda_0^2(x-1)+
  2  (x-1)\chi  \alpha^{(L)}_{0,0} \lambda_0
  \left[ {-4\lambda_0^2+3 \over (1-\lambda_0^2)^{1/2}} +
  \chi  \alpha^{(L)}_{0,0} \lambda_0 \right]\right\} +
  \nonumber\\
  &&N^{0} \Bigg\{ \left[3x-2-6 \lambda_0^2(x-1) \right] n_c
  +(x-1)\left[ (2L+1) -(2L+3) \lambda_0^2 +\left(1-\lambda_0^2 \right)
    \left(\crP_c +\anP_c \right)-4\lambda_0^2
   \left({\crc_0}^2 +2 \crc_0 \anc_0+ \anc_0^2 \right) \right]
 + \nonumber\\
 && 2 \chi (x-1) \Bigg\{ \lambda_0 (1-\lambda_0^2)^{1/2}
 \Bigg[ \sum_{\mu=-L}^{+L}   \alpha^{(L)}_{0,\mu}+
 2 \crc_\mu \anc_\mu \left[ (-1)^\mu   \alpha^{(L)}_{\mu,-\mu} +
   \alpha^{(L)}_{0,\mu} \right] +(-1)^\mu  \alpha^{(L)}_{\mu,0}
 \left( \crc_{\mu} \crc_{-\mu}+ \anc_\mu \anc_{-\mu} \right) \Bigg]
 \nonumber\\
 && -{\lambda_0^3   \alpha^{(L)}_{0,0} \over 2(1-\lambda_0^2)^{1/2}}
 \Bigg[2+3\left({\crc_0}^2 +2 \crc_0 \anc_0+  \anc_0^2 \right) +2 n_c \Bigg]
-{\lambda_0^5   \alpha^{(L)}_{0,0} \over 4(1-\lambda_0^2)^{3/2}}
\left(1+{\crc_0}^2 +2 \crc_0 \anc_0+  \anc_0^2 \right)
\Bigg\} +\nonumber\\
&&+ \chi^2 (x-1) \lambda_0^2  \Bigg\{ 1+ \sum_{\mu=-L}^{+L}
2 \crc_\mu \anc_\mu \left[ (-1)^\mu   \alpha^{(L)}_{0,0}
  \alpha^{(L)}_{\mu,-\mu} +  {\alpha^{(L)}_{0,\mu}}^2\right]+
(-1)^\mu  {\alpha^{(L)}_{0,\mu}}^2 \left( \crc_{\mu} \crc_{-\mu}+
  \anc_\mu \anc_{-\mu} \right)
\Bigg\} \Bigg\} +O(1/\sqrt{N}),\\
&& \nonumber \\
&=& H_1+H_{1/2}+H_0+O(1/\sqrt{N}),
   \label{eq:hamilHP}
\end{eqnarray}
%
%
%
%
where $ \alpha^{(L)}_{\mu,\nu}=\langle L ,\mu; L \nu | L,\mu+\nu
\rangle$ and  $\crP_c=\crc \cdot \crc=(\anP_c )^\dag$.

The term of order $N$ $(H_1)$ is exactly the mean-field energy.
Setting $\lambda=\beta/{\sqrt{1+\beta^2}}$ one gets,
%
%
%
%
\begin{equation}
\label{eq:ES}
E(N,~\beta,~x,~y)= N {\beta^2 \over (1+\beta^2)^2}
\Big[ 5x-4+ x \beta^2 + \beta  y (x-1)(4+\beta  y) \Big] ,
\end{equation}
%
%
%
%
where $y=\chi \alpha^{(L)}_{0,0}$. In the case of $L=2$, equation
(\ref{eq:ES}) reduces to the IBM ground state energy.
Note that $H_1$ only depends on $L$ through the
Clebsch-Gordan coefficient $\langle L ,\mu; L \nu | L,\mu+\nu
\rangle$ although this dependence can be absorbed in the parameter
$y$.

\begin{figure}[ht]
  \centering
  \includegraphics[width=6.5cm]{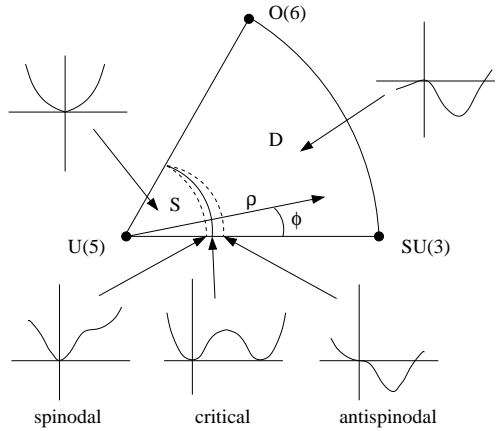}
  \caption{Qualitative phase diagram for the Hamiltonian
    (\ref{eq:hamiltonian}) and $L=2$. The insets show typical energy surfaces versus
    the deformation parameter $\beta$ in each of the phases and
    at the phase borders. The control variables of the diagram are
    defined as $\rho=1-x$ and
    $\phi=\frac{2\pi}{3\sqrt{7}}~(\frac{\sqrt{7}}{2} +\chi)$.}
\label{phaseD}
\end{figure}

$H_1$ provides the mean field energy and therefore the equilibrium
values of the order parameter. In Fig.~\ref{phaseD} is depicted the
phase diagram corresponding to $H_1$. For given parameters $x$ and $\chi
(y)$, the first step consists in
minimizing $H_1$ with respect to $\beta (\lambda)$ getting the
equilibrium value $\beta_0 (\lambda_0)$. The study of these minima
have been shown in several publications \cite{Feng81,Lopez96}, but
for completeness we summarize
here the main features:
\begin{itemize}

\item $\beta=0$ is always a stationary point. For $x<4/5$,
  $\beta=0$ is
  a maximum while for $x>4/5$ becomes a minimum. In the case of $x=4/5$,
  $\beta=0$ is an inflection point. $x=4/5$ is the point in which a minimum
  at $\beta=0$ starts to develop and defines the antispinodal line.

 \item For $\chi\neq 0 ~(y \neq 0)$ there exists a
  region, where two minima, one
  spherical and one deformed, coexist. This region is
  defined by the point where the $\beta=0$ minimum appears
  (antispinodal point) and the point where the $\beta\neq 0$ minimum
  appears (spinodal point). The spinodal line is defined by  the
  implicit equation:
\begin{equation}
\frac{3\,x}{3\,x-4}=\frac{\cal A}{\cal B}
\left (1-\left(1+\frac{\cal B}{\cal A}\right)^{\frac{3}{2}}\right )
\end{equation}
where ${\cal A}= (4 - 3\,x + 2\,(x -1)\,y^2)^2$ and 
${\cal B}=36 \,y^2 \, ( x -1)^2$. 
In the $SU(3)$ case, $\chi=-\sqrt{7}/2$, provides $x \simeq 0.820361$. 

\item In the coexistence region, the critical point is defined as the
  situation in which both minima (spherical and deformed) are
  degenerate. At the critical point the two degenerated minima are at
  $\beta_0=0$ and $\beta_0=\alpha_{0,0}^{(L)}\chi/2$ ($\beta_0=y/2$) and
  their energy is equal to zero. The critical point line can be
  calculated to be
\begin{equation}
x_c=\frac{4+y^2}{5+y^2}=\frac{4+\chi^2 \langle L ,0; L 0 | L, 0
  \rangle^2} {5+\chi^2 \langle L ,0; L 0 | L, 0 \rangle^2}.
\end{equation}
In the case of $L=2$:
\begin{equation}
x_c=\frac{4+\frac{2}{7}\chi^2 } {5+\frac{2}{7}\chi^2},
\end{equation}
being in the $SU(3)$ limit ($\chi=-\sqrt{7}/2$) $x_c=9/11$.

\item According to the previous analysis, for $\chi\neq 0~ (y\neq 0)$
  there appears a first-order phase transition, while for $\chi=0 
  (y=0)$, there is an isolated point of  second-order
  phase transition at $x=4/5$. In this last case, antispinodal,
  spinodal and critical points collapse in a single point.

\end{itemize}

The substitution of $\beta_0 (\lambda_0$) in the Hamiltonian
(\ref{eq:hamilHP}) implies that the term of order $N^{1/2}$ vanishes
because it is proportional to the derivative of $H_1$ with respect to $\lambda$.  More precisely, one has that $\partial H_1/ \partial \lambda=2 H_{1/2}$. The first quantum corrections comes from
the $N^0$ term which is a simple quadratic form in the $c$ boson
operators. It cans thus be  diagonalized through a Bogoliubov
transformation. This transformation depends on the phase, spherical or
deformed and in the next subsections both will be treated separately.

\subsection{Bogoliubov  transformation in the spherical phase}

In the spherical phase  $\beta=0$ ($\lambda_\mu=0$ for all
$\mu$) and $x>4/5$. In this case, the Hamiltonian (\ref{eq:hamilHP})
reads as,
%
\begin{equation}
  H =(3x-2 ) n_c +(x-1) \Big[ (2L+1) + \big(\crP_c +\anP_c \big) \Big]
+O(1/N),
\end{equation}
%
which is straightforwardly  diagonalized via a Bogoliubov transformation
\begin{eqnarray}
\nonumber
\crc_\mu &=& u_\mu \crx_\mu +v_\mu \tilde {\xi}_\mu\\
\tilde{c}_\mu &=& u_\mu \tilde {\xi}_\mu + v_\mu \crx_\mu
\label{eq:bogo}
\end{eqnarray}
where the coefficients verify $u_\mu^2-v_\mu^2=1$, with  $u_\mu=u_{-\mu}$
and $v_\mu=v_{-\mu}$. The phases of the coefficients are chosen so as to
minimize the mean field energy,  leading to
%
%
%
%
\begin{equation}
  H ={2L+1 \over 2} \left[-x+\Xi(x)^{1/2} \right]+
  n_\xi \: \Xi(x)^{1/2} +O(1/N),
  \label{eq:hamdiagsym}
\end{equation}
%
%
%
%
where we have introduced $\Xi(x)=x(5x-4)$ and $n_\xi $ is the number
operator for $\xi$ bosons. Note that in the spherical
phase the mean-field energy is equal to zero. In this phase, which is only
defined for $4/5\leq x \leq 1$, the spectrum is, at this order,
independent of $\chi~(y)$ and has a trivial dependence on $L$. As shown in
Ref. \cite{Vidal06_1} for $L=0$, one has to diagonalize $H$ at next
order ($1/N$) to see the role played by this parameter.

In this phase there exists a $(2L+1)$ times degenerated phonon ($5$ in
the IBM case), $\xi$. The Hamiltonian is completely harmonic and
therefore the two-phonon excitation energy is exactly twice the one-phonon excitation energy.

Another observable of interest that can be calculated easily is the
number of $L$ bosons in each state.
For the calculation of such observable the Hellmann-Feynman theorem
can be used. It establishes that the derivative of the
eigenvalue of a given operator, {\it e.g.}~the Hamiltonian,
is equal to the expectation value of the derivative of this
operator with the corresponding eigenfunction.
This leads to:
\begin{equation}
\label{eq:HF1}
\langle n_L \rangle = \frac{\partial}{\partial \theta} \Big [(1+\theta) \langle H \rangle \Big], 
\end{equation}
where $\theta=\frac{x}{1-x}$. In this case, the contribution from the mean field is zero and the
first non vanishing contribution comes from the term proportional to
$N^0$ in the energy. Therefore,
\begin{eqnarray}
\label{ndsphe1}
\langle n_L\rangle_{gs}
&=&\frac{2 L+ 1}{2} \left[ \frac{3 x-2}{\Xi(x)}-1\right]+ O(1/N), \\
\label{ndsphe2}
\langle n_L\rangle_{p\xi} &=& \langle n_L \rangle_{gs}+
p \left[\frac{3 x-2}{\Xi(x)}\right]+ O(1/N),
\end{eqnarray}
where $\langle n_L\rangle_{gs}$ stands for the expectation value of
the number of $L$ bosons in the ground state, $p$ is the
number of excited $\xi$ bosons and $\langle n_L\rangle_{p\xi}$ stands
for the expectation value of the number of $L$ bosons in the state
with $p$ excited $\xi$ bosons.
The $N^0$ correction is singular at $x=4/5$ as already noted in
similar models \cite{Dusuel05_3,Dusuel05_4}.

Note that here, we have chosen $\beta_0$ as an order parameter but,
one may have taken $\langle n_L\rangle_{gs}$ equivalently. Indeed, in
the thermodynamical limit, this quantity is only nonvanishing in the
deformed phase as we shall now see.

\subsection{Bogoliubov  transformation in the deformed phase}

In the deformed phase where $\beta_0 \neq 0$ ($\lambda_0 \neq 0$), the
situation is more complicated and strongly depends on $L$.  In the
following, we will discuss the two cases $L=0,2$ separately but we
underline that the form (\ref{eq:hamilHP}) of the expanded Hamiltonian
allows for the study of arbitrary $L$.

\subsubsection{The case $L=0$}

The case $L=0$ has recently attracted much attention because, at the
mean field level, it reproduces exactly the IBM phase diagram
(although, of course, it does not include $K=2$ excitations). In
Ref. \cite{Vidal06_1}, we have computed the finite-size corrections up
to $1/N$ order in the spherical phase. Here, we shall now treat the
deformed phase at order $(1/N)^0$. At this order, for $L=0$, the
Hamiltonian is easily diagonalized via a Bogoliubov transformation
over the $\anc$ scalar boson and one gets,
%
%
%
%
\begin{equation}
H=E(x,y,\beta_0)+ \frac{1}{2(1+\beta_0^2)}
\Big[  -x+(7x-8)\beta_0^2+2 (x-1) y \beta_0
\big(-2 -  y \beta_0+ 2\beta_0^2 \big)\Big]
+\frac{\Phi(x,y,\beta_0)^{1/2}}{2}
+ n_\xi \Phi(x,y,\beta_0)^{1/2}  +O(1/N),
\end{equation}
%
%
where $E(x,y,\beta_0)$ is given by Eq. (\ref{eq:ES}),
%
%
\begin{equation}
\Phi(x,y,\beta_0)=\frac{\Big[ x-(3x-4)\beta_0^2+2(x-1) y \beta_0 \big(2  +
y \beta_0  -\beta_0^2 \big) \Big]
\Big[ 5x-4-(19 x-20)\beta_0^2 +2(x-1) y \beta_0 \big(6  +
3 y \beta_0-7\beta_0^2-\beta_0^4\big) \Big]}{(1+\beta_0^2)^2}
\label{eq:betaL0}
\end{equation}
%
%
and $y=\chi$. 
In this case, one has a single phonon excitation with $K=0$. 
For $\beta_0=0$, one recovers expression (\ref{eq:hamdiagsym}) setting $L=0$.

Regarding the expectation value for the number of $L$ bosons, it can
be calculated as before through the Hellmann-Feynman theorem (see
Eq. (\ref{eq:HF1})).  Note that in the deformed phase there is a
contribution proportional to $N$ coming from the mean field
energy. More precisely, one has
\begin{eqnarray}
\nonumber
\label{ndL0gs}
\langle n_L\rangle_{gs} &=& N \frac{\beta_0^2}{1+\beta_0^2}+
(1-x)^2\frac{\partial}{\partial x} \left (\frac{1}{2(1+\beta_0^2)(1-x)}
\Big[  -x+(7x-8)\beta_0^2 \right .\\
&+& \left .2 (x-1) y \beta_0
\big(-2 -  y \beta_0+ 2\beta_0^2 \big)\Big]
+\frac{\Phi(x,y,\beta_0)^{1/2}}{2(1-x)}
\right )\\
\langle n_L\rangle_{p\xi} &=& \langle n_L\rangle_{gs} +
p (1-x)^2\frac{\partial}{\partial x}
\left ( \frac{\Phi(x,y,\beta_0)^{1/2}}{1-x}
\right),
\label{ndL0beta}
\end{eqnarray}
where we have used the same notation as in
Eqs. (\ref{ndsphe1},\ref{ndsphe2}) and $p$ is the number
of excited $\xi$ bosons.

\subsubsection{The case $L=2$}

In this section we will focus on the IBM case, {\it i.e.}~$L=2$.
For arbitrary $L\neq 0$, the Hamiltonian (\ref{eq:hamilHP}) must be diagonalized
for each value of $\mu$ separately. Indeed, one has
%
%
\begin{equation}
H_0=C+ \sum_{\mu =-2}^{+2} H_{\mu},
\end{equation}
%
%
where $C$ is a constant and $H_{\mu}=H_{-\mu}$.
As can be seen in Eq. (\ref{eq:hamilHP}),
$H_\mu$ depends not only on $\mu$ but also on the angular momentum $L$ via the
Clebsch-Gordan coefficients $ \alpha^{(L)}_{\mu,\nu}$.

We diagonalize separately the modes $\mu=0$, $\mu=\pm 1$, and
$\mu=\pm 2$ which correspond to the $\beta$ phonon ($K=0$), a Goldstone
phonon ($K=1$ two-fold degenerate), and the $\gamma$ phonon ($K=2$
two-fold degenerate), respectively. After the
diagonalization via a Bogoliubov transformation,
the full diagonal Hamiltonian in the deformed phase reads,
%
%
\begin{equation}
H=E(x,y,\beta_0)+ \frac{1}{2(1+\beta_0^2)}
\Big[ -5x+(19x-24) \beta_0^2+12 (x-1)y  \beta_0^3\Big]+
\sum_{\mu=-2}^{+2} \frac{\Phi_{\mu}(x,y,\beta_0)^{1/2}}{2}+
n_{\xi_\mu} \Phi_\mu^{1/2} (x,y,\beta_0)+  O(1/N),
\label{eq:hamilL2}
\end{equation}
%
%
with
{\small
%
%
\begin{eqnarray}
\Phi_0(x,y,\beta_0)&=&\frac{\Big[ x-(3x-4)\beta_0^2+2(x-1) y \beta_0
\big(2  +y \beta_0  -\beta_0^2 \big) \Big]
\Big[ 5x-4-(19 x-20)\beta_0^2 +2(x-1) y \beta_0
\big(6  +3 y \beta_0-7\beta_0^2-\beta_0^4\big) \Big]}
{\big(1+\beta_0^2 \big)^2}, \nonumber \\
 && \\
 \Phi_{\pm1}(x,y,\beta_0)&=&\frac{\Big[ x-(3x-4)\beta_0^2+(x-1) y
   \beta_0 \big(2  +y \beta_0  -2 \beta_0^2 \big) \Big]
   \Big[ 5x-4-(3 x-4)\beta_0^2 +2(x-1) y \beta_0 \big(3  +
y \beta_0-\beta_0^2\big) \Big]}{\big(1+\beta_0^2 \big)^2} ,\\
\Phi_{\pm2}(x,y,\beta_0)&=&\frac{\Big[ x-(3x-4)\beta_0^2-
2(x-1) y \beta_0 \big(2  +y \beta_0  +\beta_0^2 \big) \Big]
\Big[ 5x-4-(3 x-4)\beta_0^2 +2(x-1) y \beta_0 \big(-6  +
y \beta_0-\beta_0^2\big) \Big]}{\big(1+\beta_0^2 \big)^2},
\end{eqnarray}
%
%
}
where $y=-\sqrt{2/7}\chi$.
For $\beta_0=0$, the symmetry between modes is restored
[$\Phi_{\mu}(x,y,0)=\Xi(x)$] and one recovers the expression
(\ref{eq:hamdiagsym}) with $L=2$.

For $\beta_0 \neq 0$, the phonon excitations depend on $\mu$.
The
excitation for $\mu=0$ bosons, which corresponds to $\beta$ bosons,
is the same as in the $L=0$ case, namely
$\Phi_0(x,y,\beta_0) =\Phi(x,y,\beta_0)$.
In addition, the excitation energy for
$\mu=\pm1$ modes vanishes since for $\beta_0 \neq 0$, one has
$\Phi_{\pm 1} (x,y,\beta_0) \propto \tfrac{ \partial E(x,y,\beta)}{
\partial \beta}\big|_{\beta_0}=0$. This is in agreement with the fact
that the $\mu=\pm 1$ excitation corresponds to a rotation of the
ground state, {\it i.e.}~to a Goldstone phonon. Finally, the $\mu=\pm 2$
excitation corresponds to a $\gamma$ excitation, which is two-fold
degenerate.

For the calculation of the expectation value for the number of $d$
bosons we proceed as in the $L=0$ case and get:
\begin{eqnarray}
\nonumber
\label{ndL2gs}
\langle n_L\rangle_{gs} &=& N \frac{\beta_0^2}{1+\beta_0^2}+
(1-x)^2\frac{\partial}{\partial x} \left (
\frac{1}{2(1+\beta_0^2)(1-x)}
\Big[ -5x+(19x-24) \beta_0^2 \right .\\
&+& \left . 12 (x-1)y  \beta_0^3\Big]
+\sum_{\mu=-2}^{\mu=+2}\frac{\Phi_\mu(x,y,\beta_0)^{1/2}}{2(1-x)}
\right )\\
\langle n_L\rangle_{p\xi_\mu} &=& \langle n_L\rangle_{gs} +
p (1-x)^2\frac{\partial}{\partial x}
\left ( \frac{\Phi_\mu(x,y,\beta_0)^{1/2}}{1-x}
\right),
\label{ndL0mu}
\end{eqnarray}
where, again, the same notation as in
Eqs. (\ref{ndsphe1},\ref{ndsphe2}) is used and $p$ corresponds to the
number of $\xi_\mu$ excited bosons
($\beta$ or $\gamma$ bosons in the $L=2$ case).

%
%
\section{Numerical results}
%
%
In this section we compare the analytical results obtained in previous
sections with numerical calculations. Note that for clarity only
the firsts $0^+$ and $2^+$ states are plotted as members of the
different bands.

\subsection{The case $L=0$}
In this case, we perform the numerical calculations using the technique
presented in reference \cite{Garcia05,Vidal06_1}.
It allows to easily deal with a large number of bosons, up to a few thousands. One can reach such a number of bosons due to the underline
$O(5)$ symmetry which allows to use a seniority scheme reducing
considerably the dimension of the matrices to be diagonalized.

\begin{figure}[ht]
  \centering
  \includegraphics[width=6.5cm]{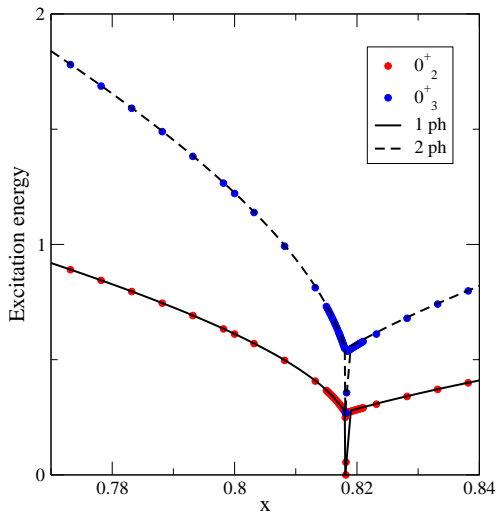}
  \caption{(Color online) Behavior of  one and two phonon excitation
    energies, in arbitrary units,  for $L=0$ as a function of
    $x$ near the critical point for $\chi=-\sqrt{7}/2$. Lines
    are the analytical results and
    dots are the numerical calculations.}
  \label{figaver1}
\end{figure}

In Fig.~\ref{figaver1} we compare the analytical
with the numerical excitation energies for a
large number of bosons, $N=5000$ (all the following calculations for
$L=0$ are performed for $N=5000$) and
$\chi=-\sqrt{7}/2$.
Note that single and two phonon
excitations are equally well described. The left part of the figure
corresponds to the deformed phase while the right part to the spherical
one. In this case there appears a first order phase transition and
at the critical point $x_c=9/11$ the ground state and the
first excited state are degenerated, one corresponding to the
spherical and the other to the deformed ground state.

\begin{figure}[ht]
  \centering
  \includegraphics[width=6.5cm]{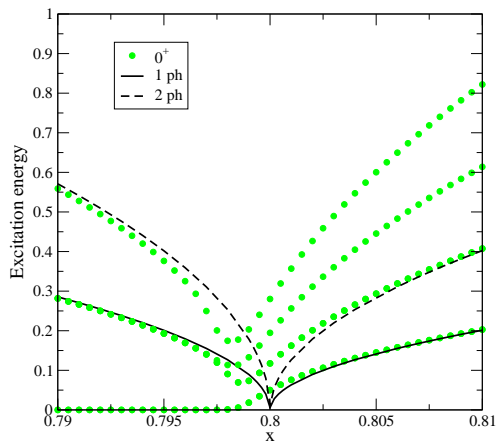}
  \caption{(Color online)
    Same as Fig.~\ref{figaver1} but for $\chi=0$. The excited
    phonon  in the
    deformed region ($x<4/5$) is the equivalent to the
    $\beta$ excitation in IBM. The lowest $0^+$ corresponds to $0^+_2$.}
  \label{figaver2}
\end{figure}

In Fig.~\ref{figaver2} we repeat the same
comparison for the case $\chi=0$. In this case a second order phase
transition appears. The energy of the first excited state becomes zero
in the deformed and in the spherical phase at the critical point,
$x_c=4/5$. At this point, regarding the analytical
calculations, the deformed $\beta$-excitation
transforms into the spherical one-phonon excitation. However,
concerning the
numerical results, the $0^+_2$ state, identified with the $\beta$ band
in the deformed sector, transforms into the two-phonon excitation
in the spherical sector.

Note that although in the spherical
phase the $N^0$ correction is independent on
$\chi~(y)$ there is a noticeable
difference between Fig.~\ref{figaver1} and Fig.~\ref{figaver2} because
for each $x$-value only the phase that corresponds to the lowest
mean field energy is plotted. The spherical phase only becomes the
most stable from $x>9/11$ on for $\chi=-\sqrt{7}/2$, while in the case
$\chi=0$ it is from $x>4/5$ on. It should be noted too that in the
deformed phase for $\chi=0$ there appear degenerate doublets of
levels due to the extra parity symmetry in the Hamiltonian in this
case.
Thus, the $\beta$ band is connected to two and three phonon excitation
in the spherical phase while the $\beta^2$ band is related to
the four and five (not shown in Fig.~\ref{figaver2}) phonon excitation
in the spherical phase.

\begin{figure}[ht]
  \centering
  \includegraphics[width=6.5cm]{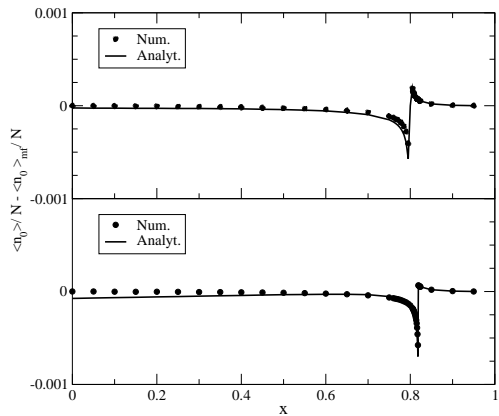}
  \caption{  
    Variation of 
    $(\langle n_{L=0}\rangle_{gs}-\langle n_{L=0}\rangle_{gs}^{mf})/N$ 
    as a function of $x$ near
    the critical point for $L=0$. Full line is for the analytical and
    circles for are numerical results. Upper figure corresponds to
    $\chi=0$ and lower to $\chi=-\sqrt{7}/2$.}
  \label{fig2}
\end{figure}

For the number of bosons,
we compare the analytical formulae with the numerical results for
the case of $\chi=-\sqrt{7}/2$ and $\chi=0$
in Fig.~\ref{fig2}. In particular we are interested in the study of
the $N^0$ corrections for the ground state,
therefore we subtract the mean field contribution, $\langle n_0
\rangle_{gs}^{mf} /N$, to both, analytical and numerical results. As
expected we observe how the $N^0$ correction improve the description
of  $\langle n_0 \rangle_{gs}$ specially near the critical point.

\subsection{The case $L=2$}

For $L=2$, the numerical calculations have been carried out
with an IBM code \cite{Isacker98} which has been modified for allowing
calculations up to $N=100$ bosons. All numerical
calculations for IBM presented below are performed for $N=100$.

For $L=2$ the case $\chi=0$ reduces to the $L=0$ situation already
discussed. In particular, the analytical ground state energy is the
same in both cases,  although the $N^0$
correction differs; there exists only one kind of excitation: the
$\beta$, while the $\mu=\pm 1$ and $\gamma$ excitations become
spurious Goldstone bosons.
The $\beta$ excitation energy is equivalent to (\ref{eq:betaL0}).
On the exact diagonalization side,
the Hamiltonian (\ref{eq:hamiltonian})
can be rewritten in terms of the generators of
an $SU(1,1)$ algebra \cite{Garcia05,Vidal06_1} in the same way that in the
$L=0$ case.
Consequently, in this section we will only consider $L=2$ with
$\chi\neq 0$. Any
$\chi$ value can be analyzed but, as an illustration, here we will present
results for the case
$\chi=-\sqrt{7}/2$ that gives the $U(5)-SU(3)$ leg in the
Casten triangle.

\begin{figure}[ht]
  \centering
  \includegraphics[width=6.5cm]{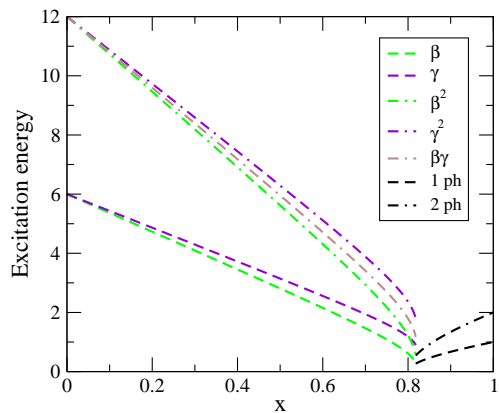}
  \caption{(Color online)
    Excitation energies (analytical), in arbitrary units,
    for one and two phonon states as a
    function of $x$ for $L=2$ and $\chi=-\sqrt{7}/2$.}
  \label{fig:analitic}
\end{figure}

First, we plot the analytical results corresponding to one and two
phonon excitations (Fig.~\ref{fig:analitic}).  In the deformed phase,
the bosons are $\beta$ and $\gamma$ excitations, while in the
spherical phase they are spherical harmonic phonons. At the critical point the
$\beta$ and the $\beta^2$ bands transform into one and two phonon
bands, respectively. However, the $\gamma$, $\beta\gamma$, and
$\gamma^2$ bands apparently disappear when entering in the spherical
phase.
Indeed, it happens because $\beta$ and $\gamma$ excitations
become degenerate for $\beta=0$. The spherical phonon excitation is a
five degenerate excitation where the deformed $\beta$ and
$\gamma$ excitations collapse together with the Goldstone boson with projection
$\pm 1$ (which is at zero energy in the deformed phase).

\begin{figure}[ht]
  \centering
  \includegraphics[width=6.5cm]{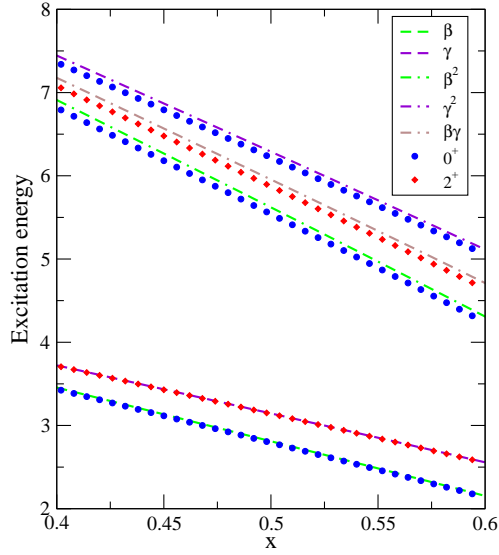}
  \caption{(Color online)
    Excitation energies (analytical and numerical), in arbitrary units,
    of one and two phonon states as a
    function of $x$ for $L=2$ and $\chi=-\sqrt{7}/2$ in the deformed
    phase. Lines correspond to analytical an
  dots to numerical results. The lowest $0^+$ state
  corresponds to $0^+_2$ and  the  $2^+$ states
  are respectively $2^+_3$ and $2^+_5$.}
  \label{fig:deformed}
\end{figure}

\begin{figure}[ht]
  \centering
  \includegraphics[width=6.5cm]{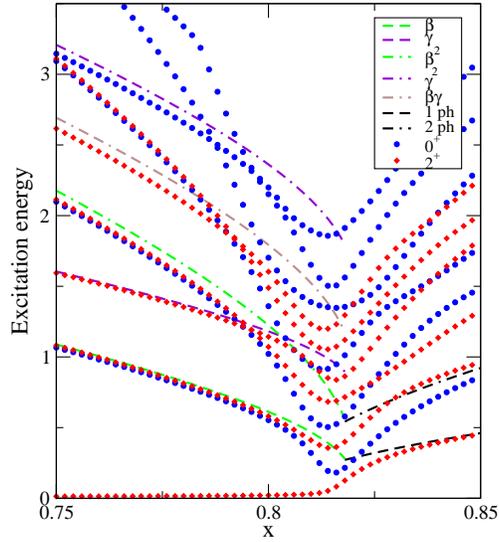}
  \caption{(Color online) Excitation energies (analytical and
    numerical), in arbitrary units,
    of one and two phonon states as a
    function of $x$ for $L=2$ and
    $\chi=-\sqrt{7}/2$ in the region around the
  critical point. The
  dots correspond to numerical results. The lowest $0^+$ state
  corresponds to $0^+_2$ and the lowest $2^+$ state
  corresponds to $2^+_1$.}
  \label{fig:crit}
\end{figure}

\begin{figure}[ht]
  \centering
  \includegraphics[width=6.5cm]{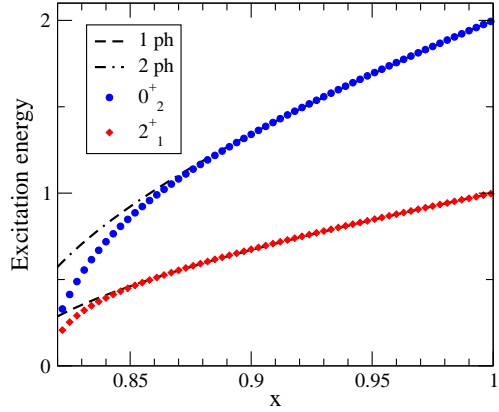}
  \caption{(Color online)
    Excitation energies of one and two phonon states, in arbitrary
    units, as a
    function of $x$ for $L=2$ and $\chi=-\sqrt{7}/2$.
    Lines correspond to analytical an
    dots to numerical results.}
  \label{fig:spherical}
\end{figure}

In order to compare analytical and numerical results
we will split the analysis
in three different regions: deformed phase (Fig.~\ref{fig:deformed}),
critical region (Fig.~\ref{fig:crit}) and spherical phase
(Fig.~\ref{fig:spherical}).  The harmonic character of the results is
observed in all these plots.
\begin{figure}[ht]
  \centering
  \includegraphics[width=6.5cm]{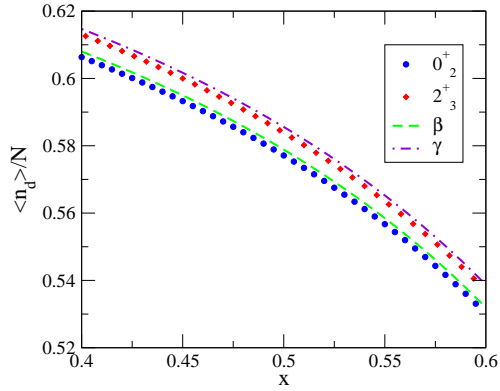}
  \caption{(Color online)
    $\langle n_d \rangle/N$ as a function of $x$ for $L=2$ and
    $\chi=-\sqrt{7}/2$,
  in the deformed phase for one phonon states. Lines correspond to
  analytical  and dots to numerical results.}
  \label{fig:deformednd}
\end{figure}

\subsubsection{Deformed phase}
In the deformed phase (Fig.~\ref{fig:deformed})
one and two phonon excitations are
clearly separated in energy. Note that the excitation energy for the $\gamma$
band is higher than the corresponding one for the $\beta$ band, although
for $x=0$ ($SU(3)$ limit) they are degenerated. Also note that the
$\gamma^2$ excitation carries the angular momentum projections $K=0,
\pm 4$ which in this approach are degenerated.
\begin{figure}[ht]
  \centering
  \includegraphics[width=6.5cm]{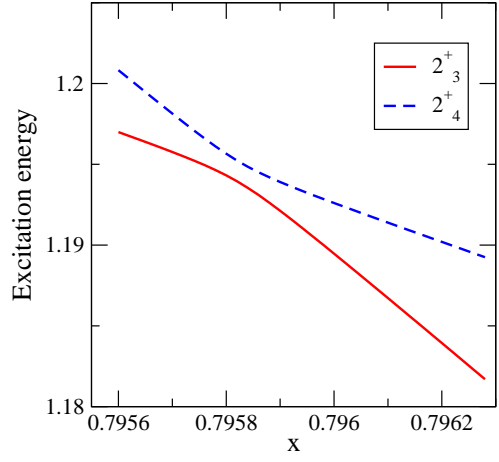}
  \caption{(Color online)
    Excitation energies (numerical) of $2^+_3$ and $2^+_4$ states  at
    the region of closest approach (see text) for $L=2$ and
    $\chi=-\sqrt{7}/2$.}
  \label{fig:zoomL2}
\end{figure}

The correspondence between numerical and analytical states
is as follows: $\beta$ band is identified with $0^+_2$, $\gamma$ with
$2^+_3$, $\beta^2$ with $0^+_3$, $\beta\gamma$ with $2^+_5$ and
$\gamma^2$ with $0^+_4$. Note that the state $2^+_2$ belongs to the
$\beta$ band, while $2^+_4$ to the $\beta^2$ band.

The overall agreement between analytical and numerical results is
satisfactory and improves the description given in \cite{Garcia98}
for single and double phonon
excitations, although in the present approach, no mixing between the different kind of excitations appear.

The average number of $d$ bosons in the deformed phase,
normalized to the total number of
bosons,  is depicted in
Fig.~\ref{fig:deformednd} for one phonon states (the results for two
phonon states are not presented for clarity).
It can be observed a  smooth decrease of $\langle n_d \rangle$
when $x$ is increasing
as it is expected when approaching the spherical phase.
\begin{figure}[ht]
  \centering
  \includegraphics[width=6.5cm]{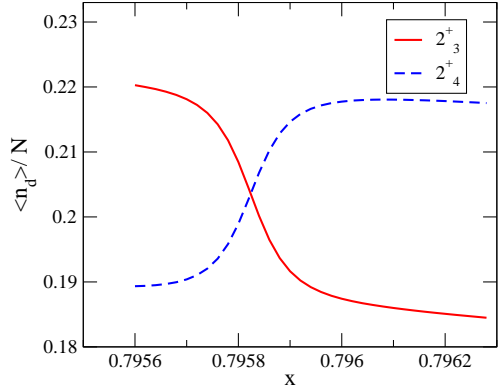}
  \caption{(Color online)
    $\langle n_d \rangle/N$ (numerical) of $2^+_3$
    and $2^+_4$ states  at
    the region of closest approach (see text) for $L=2$ and
    $\chi=-\sqrt{7}/2$.}
  \label{fig:zoomL2nd}
\end{figure}

\subsubsection{Critical area}
The comparison around the critical area (Fig.~\ref{fig:crit}) becomes
complicated because one and two phonon states have comparable energies
and there appears interchange of character between states. For example
at the critical point, the $\beta^2$  is at lower energy than the
$\gamma$ excitation.

Starting at $x=0.75$ the correspondence between analytical and numerical states is
similar  to the one given in preceding section,
but already at $x=0.8$ different states interchange its
character. The correspondence between states is presented in table
\ref{tab:correspondence}. From this table it is clear that there
exists an
interchange of character between the states corresponding to the
$\gamma$, $\beta^2$, $\beta\gamma$, and $\gamma^2$ bands.

\begin{table}
\begin{tabular}{|l|l|l|l|}
\hline
 &  ~~$x=0.75$ ~~ & ~~$x=9/11$~~  &  ~~$x=0.85$ ~~ \\
\hline
 ~~$\beta$        &  ~~$0^+_2$,$2^+_2$ & ~~$0^+_2$,$2^+_2$&\\
 ~~$\gamma$       &  ~~$2^+_3$         & ~~$2^+_4$        &\\
 ~~$\beta^2$      &  ~~$0^+_3$,$2^+_4$ & ~~$0^+_2$,$2^+_3$& \\
 ~~$\beta\gamma$  &  ~~$2^+_5$         & ~~$2^+_6$        &\\
 ~~$\gamma^2$     &  ~~$0^+_5$         & ~~$0^+_7$        &\\
 ~~$1$ phonon     &  ~~               &                & ~~$2^+_1$\\
 ~~$2$ phonons    &  ~~               &                & ~~$0^+_2$,$2^+_2$\\
\hline
\end{tabular}
\caption{Correspondence between analytical and numerical states for
  three values of $x$: $x=0.75$ deformed phase, $x=9/11$ critical
  point and $0.75$ spherical phase. Only $0^+$ and $2^+$ states are
  indicated explicitly.}
\label{tab:correspondence}
\end{table}
\begin{figure}[ht]
  \centering
  \includegraphics[width=8.5cm]{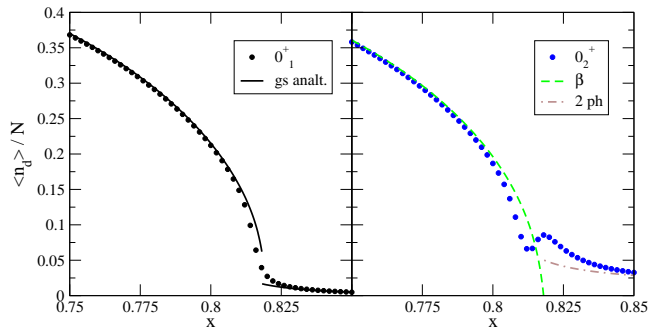}
  \caption{(Color online)
    $\langle n_d \rangle/N$
    as a function of $x$, for $L=2$ and $\chi=-\sqrt{7}/2$,
    at the critical area for one phonon states. Lines correspond to
    analytical  and dots to numerical results.}
  \label{fig:critndgs}
\end{figure}

An interesting question that arises is if the interchange of character
is due either to level crossing or to level repulsion. We have to take into
account that the transition between $SU(3)$ and $U(5)$ is not an
integrable path \cite{Arias03_2}, {\it i.e.}~a complete
set of mutually commuting Hermitian operators does not exist.
This implies that
crossings are forbidden and only repulsion is allowed. In particular, in
the thermodynamical limit the repulsion becomes anti-crossing, {\it
  i.e.}~infinity repulsion. In Fig.~\ref{fig:zoomL2} we show a zoom
of one apparent crossing in
Fig.~\ref{fig:crit} between $2^+$ states in the region around
$x=0.796$, it is clearly seen that the levels indeed repel each other
as expected.
In order to illustrate this result and show how the two involved
levels  interchange
their character, we present  in Fig.~\ref{fig:zoomL2nd}
the expectation value of the $d$ boson number in both
states. It is clearly observed that the states interchange their
properties at the point  of closest approach.

The average number of $d$ bosons, normalized to the total number of
bosons, in the region around the critical point is depicted in
Fig.~\ref{fig:critndgs} for the ground state (left panel) and for the
$\beta$ band (right panel).
One important feature is the discontinuity appearing at $x_c=9/11$
due to the existence of a first order phase transition. In the
evolution of the $\beta$ band, it appears a kink in the numerical
results at the critical point. This behavior at the critical point
has been already observed for other observables such as isomer shifts
\cite{Iachello04}, derivative of the ratios of $4_1^+ / 2^+_1$
excitation energies \cite{Werner02} or
$B(E2:4^+_1\rightarrow 2^+_1)/B(E2:2^+_1\rightarrow 0^+_1)$ \cite{Rosensteel05}.
Also note that the $0^+_2$ state
transforms into a two phonon band when passing to the spherical
phase.
\begin{figure}[ht]
  \centering
  \includegraphics[width=6.5cm]{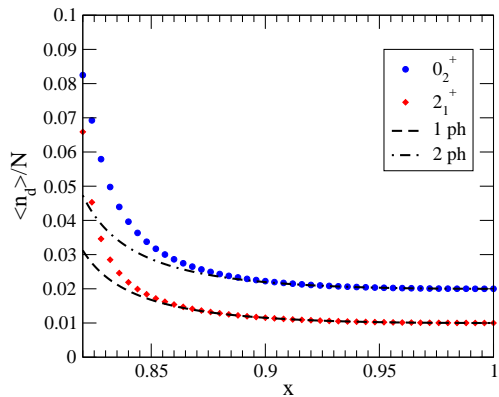}
  \caption{(Color online)
     $\langle n_d \rangle/N$ as a function of $x$ for $L=2$ and
     $\chi=-\sqrt{7}/2$,
    in the spherical phase for one and two phonon states.
    Lines correspond to analytical  and dots to numerical results.}
  \label{fig:sphericalnd}
\end{figure}

\subsubsection{Spherical phase}
The last region of interest is the spherical phase
(Fig.~\ref{fig:spherical}). Here, there exist a five degenerated
phonon excitation. The correspondence between the analytical and
the numerical
results is clear: one phonon excitation corresponds to the state
$2_1^+$ while two phonon excitation to the state $0_2^+$ (also to
$2_2^+$ and $4_1^+$ states).

The average number of $d$ bosons, normalized to the total number of
bosons, in the spherical region is depicted in Fig.~\ref{fig:sphericalnd}
for the one and two phonon states.
The main discrepancies between numerical and analytical results, as
expected, appear close to the critical point. Note that the structure of
the states is very simple and already for $x=0.9$ the number of $d$
bosons is fixed to $1$ and $2$ for one and two phonon states,
respectively.

\section{Summary and conclusions}

In this paper we have studied two--level boson models characterized by
a lowest scalar $s$-boson and an excited $L$ boson through a
Holstein-Primakoff transformation that allows to treat explicitly
order by order a $N$ expansion. This treatment shows that only the
leading $N$ term of the ground state energy is correct in a mean field
(or Hartree-Bose) approach. We stress that the equilibrium nuclear
shape corresponding to an IBM Hamiltonian should be obtained only
considering the leading $N$ term of the ground state energy.

Depending on the
value of $L$, models of interest in different fields can be
obtained. Thus, $L=0$ is related to the Lipkin model first introduced
in Nuclear Physics and then used in many fields, $L=1$ is the
vibron model of interest in Molecular Physics, $L=2$ is the
Interacting Boson Model of nuclear structure, etc. We have presented
a method to go accurately beyond the standard mean field treatment so
as to be able to compute finite size corrections to several spectroscopic
observables. The model Hamiltonian used is a generalization for
arbitrary $L$ of the Consistent Q Hamiltonian in the IBM. This
Hamiltonian depends on two control parameters and changes in them
allow to explore the full model space and the corresponding phase
diagram. Although the formalism is general for any $L$ value, we have
concentrated in the cases $L=2$ (IBM) and $L=0$. Both spherical and
deformed phases have been studied with special emphasis in the gap for
single and double excitations and the expectation values of the number
of $L$ bosons in different states. Analytic results have
been validated by comparison with full numerical calculations.


\acknowledgments
This work has been partially supported by the Spanish Ministerio
de Educaci\'on y Ciencia
and by the European regional development fund (FEDER) under
projects number BFM2003-05316-C02-02, FIS2005-01105, and
FPA2003-05958.



\end{document}